# Flux pinning in superconducting multilayer 2H-NbSe$_2$ nano-step junction

Minseong Kwon[1], Mingi Kim[1], Yoonji Gong[2], Heeyeon Lee[1] and Young Duck Kim[1,2,3*]

[1] Department of Physics, Kyung Hee University, Seoul, 02447, Republic of Korea

[2] KHU-KIST Department of Converging Science and Technology, Kyung Hee University, Seoul, 02447, Republic of Korea

[3] Department of Information Display, Kyung Hee University, Seoul, 02447, Republic of Korea

* Corresponding author: ydk@khu.ac.kr

**Abstract**

Superconductors exhibit dissipationless supercurrents even under finite bias and magnetic field conditions, provided these remain below the critical values. However, type-II superconductors in the flux flow regime display Ohmic dissipation arising from vortex dynamics under finite magnetic fields. The interplay between supercurrent and Ohmic dissipation in a type-II superconductor is dictated by vortex motion and the robustness of vortex pinning forces. In this study, we present an experimental investigation of the superconducting phase transitions and vortex dynamics in the atomically thin type-II superconductor 2H-NbSe$_2$. We fabricated a high-quality multilayer 2H-NbSe$_2$ with a step junction, demonstrating supercurrent in clean limit below a critical temperature of 6.6 K and a high residual resistance ratio of 17. The upper critical field was estimated to be 4.5 T and the Ginzburg-Landau coherence length 8.6 nm. Additionally, we observed phase transitions induced by vortex viscous dynamics in the 2H-NbSe$_2$ step junction. Analysis of the pinning force density using the Dew-Hughes model indicates that the pinning force in the 2H-NbSe$_2$ device can be attributed to step junction, related to the surface-$\Delta\kappa$ type of pinning centers. Our findings pave the way for engineering pinning forces by introducing artificial pinning centers through partial atomic thickness variation in layered 2D superconductors while minimizing unwanted quality degradation in the system.

# 1. Introduction

Atomically thin single-crystal superconductor 2H-NbSe$_2$ enables the fabrication of the ultraclean superconducting devices and the characterization of superconductivity down to the monolayer limit. Furthermore, the strong proximity effect of 2H-NbSe$_2$ with atomically clean van der Waals interfaces between other two-dimensional (2D) materials and bulk superconductors paves the way for innovative device architecture and high-performance quantum devices such as spin-charge conversion devices [1, 2], Transmon qubits [3], and Josephson junction devices [4-8]. Additionally, the layer-dependent superconducting properties of 2H-NbSe$_2$ allow the manipulation of critical temperature ($T_c$), critical field ($H_c$), critical current density, and BCS superconducting gap [9-11], facilitating fundamental studies of the Berezinskii-Kosterliitz-Thouless (BKT) transition in intrinsic 2D systems [10-13], where $d < \xi$ ($d$ denotes the thickness of 2H-NbSe$_2$, and $\xi$ is the Ginzburg-Landau coherence length).

Type-II superconductors exhibit superconductivity, such as the absence of electrical resistance and the expulsion of a magnetic field (Meissner effect) below the critical temperature and fields. Typically, type-II superconductors possess higher critical fields without losing their superconducting properties compared to type-I superconductors. These features originate from the existence of vortices, which are quantized magnetic fluxes penetrating the material under an applied external magnetic field. Thus, type-II superconductors exhibit a nonequilibrium phase of mixed superconducting and normal properties under applied external magnetic fields. In particular, flux flow phases are highly sensitive to the motion of vortices in type-II superconductors and the formation of 2D flux lattices.

In this study, we investigated the type-II superconductivity phases of 2H-NbSe$_2$ under the influence of a magnetic field and current densities that generate fluxons through vortex supercurrents [14]. Within the type-II superconducting flux flow phase, the Ohmic dissipation transport under applied external current densities originates from the dynamics of vortices, influenced by the current densities and flux pinning due to defects, impurities, and inhomogeneity of superconductor materials [15]. We chose single-crystal multilayer 2H-NbSe$_2$ as the ultraclean type-II superconductor to reduce undesired defects and impurities-related flux pinning effects. The significant suppression of random flux pinning centers in 2H-NbSe$_2$ allows exploration of the intrinsic dynamics of vortices under the flux flow phase and Ohmic transport mechanisms, viscous flows, and flux pinning-induced Lorentz force at the step junction. This study provides a deep understanding of flux flow phases in 2D limits and the engineering of quantum device applications.

## 2. Results and Discussion

Niobium diselenide (NbSe$_2$) is a part of the layered transition metal dichalcogenides (TMDCs) family and has different polytypes such as hexagonal (2H) and trigonal (3R) crystal structures. In this study, we used single-crystal 2H-NbSe$_2$, which is arranged in a hexagonal lattice with Se-Nb-Se layers strongly bound together, while each layer is held together by relatively weak van der Waals forces, as shown in Fig. 1(a). The weak van der Waals forces between layers make 2H-NbSe$_2$ suitable for mechanical exfoliation down to monolayers from bulk single crystals. Due to its single crystalline structure and layered material properties, we employed 2H-NbSe$_2$ as the atomically thin, ultraclean type-II superconductor to minimize the defect and impurities-induced flux pinning effect.

To investigate the intrinsic type-II superconductivity and phase transitions of multilayer 2H-NbSe$_2$, we realized a multi-terminal transport device as shown in Fig. 1b. Details regarding the device fabrication and electrical measurements can be found in Methods. Atomic force microscopy reveals the thicknesses of the step junction structure to be 33.5 nm (30 layers) for the thicker region and 25.7 nm (23 layers) for the thinner region as shown in Fig. 1c. Figure 1d shows the temperature dependence of the four-probe resistances of the 2H-NbSe$_2$ step junction with 0.4 mA of current bias. We observe the sharp superconducting phase transition and zero resistance state at low temperatures. We define the critical temperature $T_c$ as the corresponding temperature where the resistance ($R$) is decreased to 90% of its normal state resistance ($R_N$) right above the onset of the superconductivity. Because 2H-NbSe$_2$ exhibits varying $T_c$ depending on its thickness [10,11], we observe a series of superconducting phase transitions as illustrated in the inset of Fig. 1d. From our four-probe measurements of the 2H-NbSe$_2$ step junction device under zero magnetic field, $T_c$ of the thicker region (33.5 nm) is determined to be 6.64 K, and the $T_c$ of the thinner region (25.7 nm) is found to be 6.52 K. We find the residual resistance ratio ($R_{rr} = R(300\ K)/R(7\ K)$) of the device to be about 17. Since $R_{rr}$ is a quantity that strongly depends on the amounts of crystallographic defects, these results are consistent with the superconducting properties of high-quality 2H-NbSe$_2$ [9, 10, 16].

We investigated the temperature-dependent resistance under the external out-of-plane magnetic fields ($B$) and the current-voltage ($IV$) characteristics across various temperatures to further characterize the superconducting phase transition. Figure 2a shows the normalized resistance ($R/R_N$) as a function of temperatures from 1.5 K to 10 K under various $B$ ranging between 0.5 T to 3 T. These observations enable us to determine the temperature-dependent upper critical fields $H_{c2}(T)$, where $H_{c2}$ is defined as $B$ when $R = 0.9R_N(B)$ for a specified temperature. Here, we can estimate only a single $H_{c2}(T)$ value. This can be attributed to the marginal difference between the $H_{c2}(T)$ values corresponding to the thicker and thinner regions. Figure 2b shows experimentally estimated $H_{c2}(T)$ as a function of temperature (red dots), and the estimated $H_{c2}(T)$ of 2H-NbSe$_2$ is consistent with the theoretical linearized Ginzburg-Landau relation (black line) [17]. According to the linearized Ginzburg-Landau theory, $H_{c2}(T)$ is described as follows $H_{c2}(T) = H_{c2}(0)(1- T/T_c)$, where $H_{c2}(0)$ is the upper critical field at the absolute zero. Here, we take $T_c$ as $T_{c,\ avg}$, the average critical temperature of the two regions (6.58

K), as it aligns with our approach of a singly estimated $H_{c2}(T)$. The value of $H_{c2}(0)$ is given by $H_{c2}(0) = \Phi_0/2\pi\xi^2(0)$, where $\Phi_0$ is the magnetic flux quantum and $\xi$ is the Ginzburg-Landau coherence length. From the linearized Ginzburg-Landau theory analyses for upper critical fields, we can estimate $\xi(0)$ of 8.6 nm for the 2H-NbSe$_2$ step junction device. With $R_N$, we calculate the 2D mean free path, $l = h/e^2(\sigma_{2D}/(2\pi n_{2D})^{1/2})$, which can judge the device's limit of disorder by comparing with $\xi(0)$ [18]. Here, $h$ is the Planck constant, $e$ is the electron charge, $\sigma_{2D}$ is two-dimensional conductivity ($\sigma_{2D} = (L/W)(N/R_N)$ where $L$ and $W$ are the length and width of the device, and $N$ is the number of 2H-NbSe$_2$ layers), and $n_{2D}$ is carrier concentration. For the thicker region, which constitutes the majority of the device (95 % in area), the physical quantities are $L = 2.5$ μm, $W = 27.9$ μm, $N = 30$, and $R_N = 0.31$ Ω. Since the variation in $n_{2D}$ with thickness is minimal, we use the $n_{2D}$ of bulk to calculate $l$ [10, 18, 19]. We find $l = 34$ nm, and the device is in the clean limit, $\xi < l$.

Figures 2c and 2d display symmetric $IV$ characteristics at 2 K and 6.6 K and calculated differential resistances (d$V$/d$I$). In Fig 2d, green and red arrows indicate two distinctive critical current peaks at 1.2 mA and 1.8 mA in d$V$/d$I$ at 2 K. These peaks are observed in step junction devices and attributed to the different critical currents of 2H-NbSe$_2$ depending on its thickness [9]. As shown in Figure 2e, we observe critical behavior as the temperature is increased up to 7 K. The dark blue region in Fig. 2e clearly indicates a zero-resistance state even near the $T_c$. Figure 2f illustrates similar trends of two critical current peaks regarding the reduced temperature ($T/T_c$) and the normalized critical current densities ($j$, a ratio of the critical current density at a specific temperature to the critical current density at 1.5 K). The fundamental depairing current density at zero temperature in the clean limit for type-II superconductors, $J_{depair}(0)$, can be calculated using the formula $J_{depair}(0) = 0.826(H_{c0}/\lambda_0)$, where $H_{c0}$ is the zero-temperature thermodynamic critical field and $\lambda_0$ is the BCS zero-temperature London depth [20-23]. The detailed calculation is provided in Supplementary Material. For our device's thicker region, the calculated $J_{depair}(0)$ is $4\times10^7$ A/cm$^2$, which is two order of magnitude higher than the critical current density at 1.5 K ($2\times10^5$ A/cm$^2$). This discrepancy can be attributed to the ideal conditions assumed in the calculation. Introducing the finite Dynes broadening parameter of the device and a non-zero impurity scattering rate into the analysis could help narrow this gap. This aspect could be explored further in future studies to refine the understanding of the critical current density under realistic conditions.

In the equilibrium state, type-II superconductors under an external magnetic field form a vortex lattice to minimize the free energy, as shown in the left panel of Fig. 3a. In the nonequilibrium state, Lorentz force due to the applied current density under a magnetic field will be competing with pinning forces ($F_p$) at microscopic vortex lattice pinning centers of the type-II superconductors. The right panel of Fig. 3a shows a viscous flow phase where vortices can move freely when Lorentz force becomes larger than $F_p$. Electric fields are held in the type-II superconductor by the motion of vortices, hence dissipative charge transport. Therefore, the pinning force opposing the Lorentz force is crucial for applications [24-26].

We estimated the pinning current density ($J_p$) at the intersection between the tangent line for values lower than $E = 10$ V/m under various external applied magnetic fields ranging from 0 T to 3.5 T, with a 0.1 T step at 1.5 K, as shown in the inset of Fig 3b. $J_p$ represents the maximum current density for the dissipation-free charge transfer phase due to the dominant flux pinning in 2H-NbSe$_2$. The Ohmic dissipation transport behaviors are observed except for the current density below $J_p$. Thus, with $J_p$, we can define flux pinning force density with Lorentz force density as $F_p = J_p \times B$. Figure 3c elucidates the dependency of normalized flux pinning force density ($f_p = F_p/F_p^{max}$) on the normalized magnetic field ($h = B/B_p^{max}$, where $B_p^{max}$ is the maximum $B$ that flux pinning presents). Here, we use $B_p^{max} = 3.67$ T, practically assumed based on a linear approximation of $J_p$ above 3 T.

The dominant flux pinning mechanism in the 2H-NbSe$_2$ step junction device can be identified through the Dew-Hughes scaling law analysis of $f_p$ as a function $h$ [27-29]. The $f_p$ of the device reaches its maximum at $h = 0.57$ ($B = 2.1$ T), which is differ from devices without a step junction and closely matches the maximum position for a $\Delta\kappa$-type surface pinning center (see Supplementary Material Table S1) [30]. To further analysis, we performed a fitting procedure. In our fitting, we considered all possible terms in the Dew-Hughes model, represented by the equation $\sum C_i h^p (1-h)^q$, where $p$ and $q$ are the exponents characteristics of different pinning mechanisms in the Dew-Hughes model. We treated $C_i$, the coefficient representing the contributions of each class of pinning centers, as the fitting parameters. This approach allowed us to quantify the relative contributions of various pinning mechanisms to the flux pinning behavior of the step junction device. We find the $\Delta\kappa$-type surface pinning center mainly contributes to the fitting, while the contributions from the $\Delta\kappa$-type volume, normal-type surface, and $\Delta\kappa$-type point pinning centers are negligibly small (see Supplementary Material Table S1). This suggests that the $\Delta\kappa$-type volume, normal-type surface, and $\Delta\kappa$-type point pinning centers play a minimal role in the overall flux pinning behavior of the 2H-NbSe$_2$ step junction device. Consequently, we refined our fitting using the equation $\sum C_i h^p (1-h)^q$ for the two main pinning centers: $\Delta\kappa$-type surface and normal-type point pinning centers. The final fitting yielded $C_i$ values of 1.44 for the normal-type point pinning center and 4.56 for the $\Delta\kappa$-type surface pinning center. The normal-type point pinning center can be attributed to selenium vacancies in the crystal, and the $\Delta\kappa$-type surface pinning center can be attributed to the step junction in the device.

When the applied current just exceeds $I_p$, vortex motion occurs, resulting in dissipation. This vortex motion can be described as simple viscous flow $E/J = \Phi_0 B/\eta$, where $\Phi_0$ is the superconducting magnetic flux quantum ($\Phi_0 = h/2|e|$. Here, $h$ is the Planck constant, and $e$ is the electron charge) and $\eta$ is a viscosity coefficient [31]. As shown in Fig 4a, dissipation is present even below the critical current except under zero $B$ and increases proportionally with the magnetic field. These dissipations due to viscous flow are also shown in Fig 4b as a moderate increase of $R/R_N$ proportional to the magnetic field. Due to the superconductor-normal metal transition, the viscous flow is followed by a rapid increase of $R/R_N$ with increasing $B$, and $R/R_N$ saturates above 0.9.

To investigate the superconducting phases in 2H-NbSe$_2$ step junction more clearly, we present a color plot of the $B$ derivative values of $dV/dI$ as shown in Fig 4c. Below $I_p(B)$, indicated by a dotted line in Fig 4c, vortices are pinned, resulting in zero derivative values. Due to the dissipation by the viscous vortex motion, non-zero derivative values are observed to correspond to a value proportional to $1/\eta$ between $I_p(B)$ and the dash-dot line in Fig 4c. Beyond the dash-dot line, the sequential superconductor-normal metal transition occurs, influenced by the step junction. The transition begins from the thinner region at its critical current density of the thinner region ($J_{Thin}$), and beyond the critical current density of the thicker region ($J_{Thick}$), the entire device transitions to the normal metal state. These critical behaviors of the 2H-NbSe$_2$ step junction device are summarized in Figure 4d as a phase diagram.

## 3. Summary

We fabricated a superconducting multilayer 2H-NbSe$_2$ step junction device that exhibits nonlinear superconducting phases at a finite temperature and current under a perpendicular magnetic field. The basic critical parameters are determined to be $T_c$ = 6.64 K for the thicker region and 6.52 K for the thinner region, $H_{c2}(0)$ = 4.5 T, and $\xi(0)$ = 8.6 nm with an applied current of 0.4 mA. The high-quality nature of the device is substantiated by achieving a $R_{rr}$ of 17. We observe sequential superconductor-normal metal transitions due to the step junction's internal difference in $T_c$ and the critical currents with the normalized current densities showing similar temperature dependence. At 1.5

K, we observe nonequilibrium behaviors of the device related to the flux pinning and the viscous vortex motion. Pinning force density analysis using the Dew-Hughes model reveals that the pinning force in the device can be attributed to both surface-Δκ and point-normal types of pinning centers. Our observations suggest that atomic thickness engineering of 2D superconductors could effectively tune the vortex pinning physics in a 2D van der Waals superconductor without introducing unwanted disorders in the system. This study suggests that this pinning mechanism could provide a new way to enhance the functionality of 2D superconductors for enhanced critical current and high-quality superconductor qubits.

## Methods

### Device fabrication

We first exfoliated single-crystalline multilayer 2H-NbSe$_2$ flakes on a 285 nm thick SiO$_2$/Si substrate from bulk crystal by standard mechanical exfoliation method [32]. To minimize the degradation of crystal quality due to oxidation and to achieve Ohmic contact, we pursue the exfoliation of 2H-NbSe$_2$ in a nitrogen-filled glove box (Labstar130, Mbraun) with less than 0.1 ppm oxygen and water. Using an optical microscope inside the glove box, we selected a multilayer crystal flake with a stepped edge based on the optical contrast of each crystal. Electron beam resists (950 PMMA A6, MicroChem) were spin-coated onto the sample. A PMMA-based deposition mask for the electrodes was patterned on the sample using electron beam lithography with a scanning electron microscope (Sigma300, Zeiss) equipped with the nanometer pattern generation system (NPGS, JC Nabity Lithography Systems) in the Multi-dimensional Material Research Center at Kyung Hee University. Subsequently, Cr/Pd/Au (2 nm/20 nm/40 nm) were deposited using an electron beam evaporator. The sample fabrication process was completed with a lift-off procedure, immersing the device in acetone overnight to remove the electron beam resists and residual metal films. We then characterized the topography and thickness of the 2H-NbSe$_2$ step junction device using atomic force microscopy (XE-100 of Park Systems) in contact mode.

### Low temperature transport measurement

To characterize the superconductivity and phase transition of the multilayer 2H-NbSe$_2$ device, we loaded the sample into a cryostat (TeslatronPT, Oxford Instruments) in the Multi-dimensional Material Research Center at Kyung Hee University. The cryostat allowed us to cool the sample down to 1.5 K and apply a magnetic field up to $B$ = 12T. We performed DC four-point probe measurements as depicted in Fig. 1b with systematically varying temperatures and the applied external out-of-plane magnetic field.

**Acknowledgements**

The authors thank Chaun Jang and Myunglae Jo for the discussions. This work was supported by a grant from Kyung Hee University (KHU-20192441).

**Author contribution**

M. Kwon and Y.D.K. designed the experiment. M. Kwon, M. Kim, and Y.G. fabricated the device. M. Kwon, M. Kim, and H.L. performed the measurement. M. Kwon and Y.D.K. analyzed the data and wrote the paper. All authors contributed to the discussion and comments on the manuscript.

**Declaration of competing interests**

The authors declare no competing financial interests.

**Data availability**

The datasets generated during the current study are available from the corresponding author upon reasonable request.

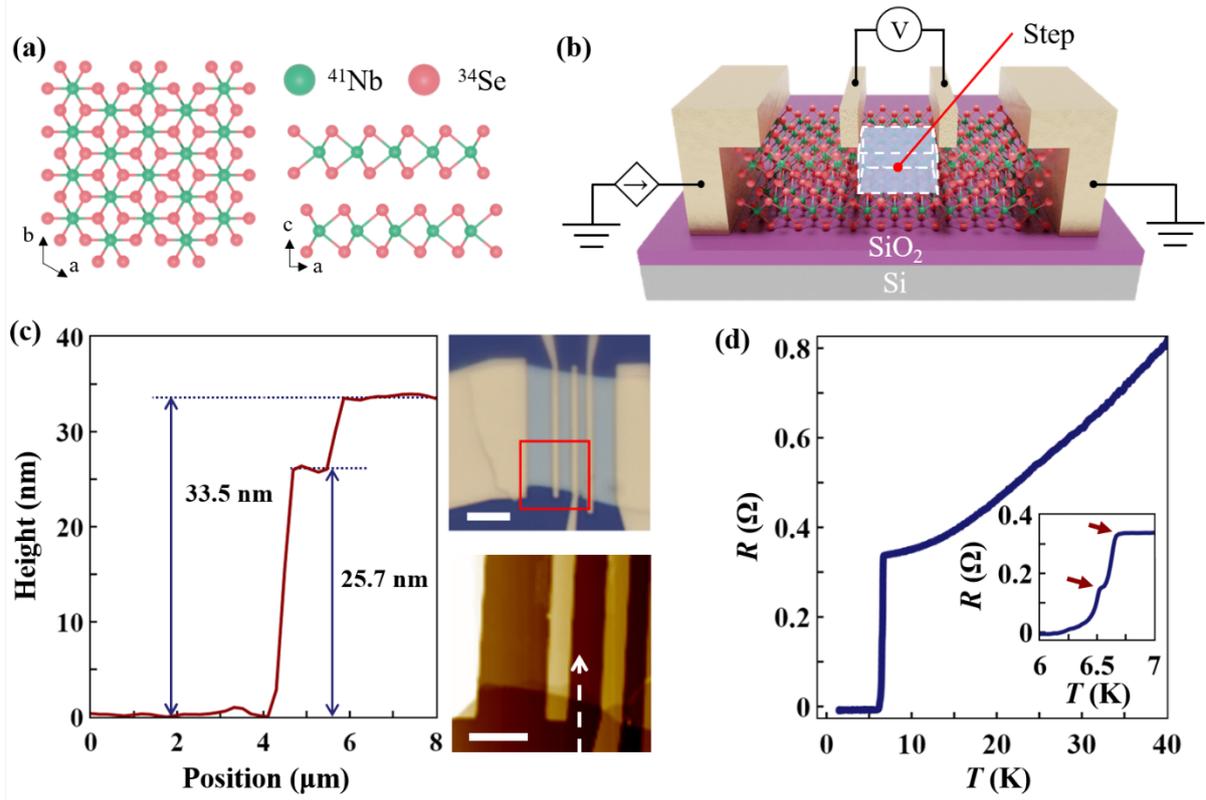

Figure 1. Temperature-dependent resistance of NbSe$_2$ device (a) Crystal structure of 2H-NbSe$_2$. (b) A schematic image of device and DC four-probe measurements. (c) The image of the device. Left: thickness profile along the white arrow on the lower right panel, the AFM topography. Upper right: the optical image of the multilayer NbSe$_2$ device. The length of the scale bar is 5 μm. Lower Right: The AFM topography of the red boxed area in the optical image. (d) Superconducting transition in NbSe$_2$. The critical temperatures of the device are 6.64 K (33.5 nm region) and 6.52 K (25.7 nm region) under applied current of 0.4 mA and zero magnetic field. (Inset: cropped R-T from 6 K to 7 K, Each arrow indicates the superconducting transition of 33.5 nm region and 25.7 nm region)

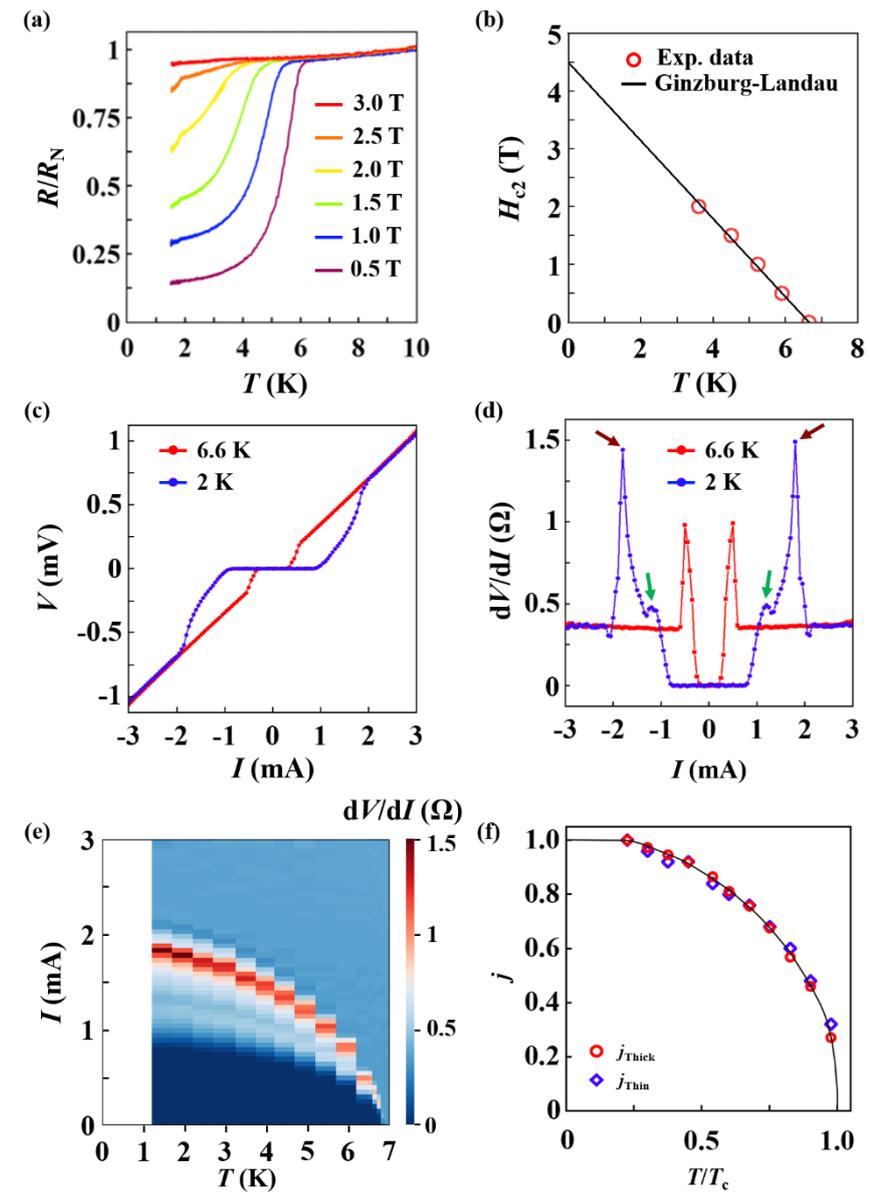

Figure 2. Superconducting phase transitions and the temperature dependence of critical parameters. (a) Temperature-dependent $R/R_N$ under various $B$ and $I$ = 0.4 mA. (b) Upper critical field ($H_{c2}$) versus temperature. Red dots indicate measured $H_{c2}$ at a specific temperature. The black line shows the linearized Ginzburg-Landau fitting. (c) Current-voltage characteristics at 2 K and 6.6 K. (d) Calculated differential resistances (d$V$/d$I$) at 2 K and 6.6 K. Green and red arrows indicate critical current peaks at 2 K. (e) Colourmap of d$V$/d$I$ under varying $T$ and $I$. (f) Normalized current density vs. reduced temperature. The solid line shows a guide for the eye.

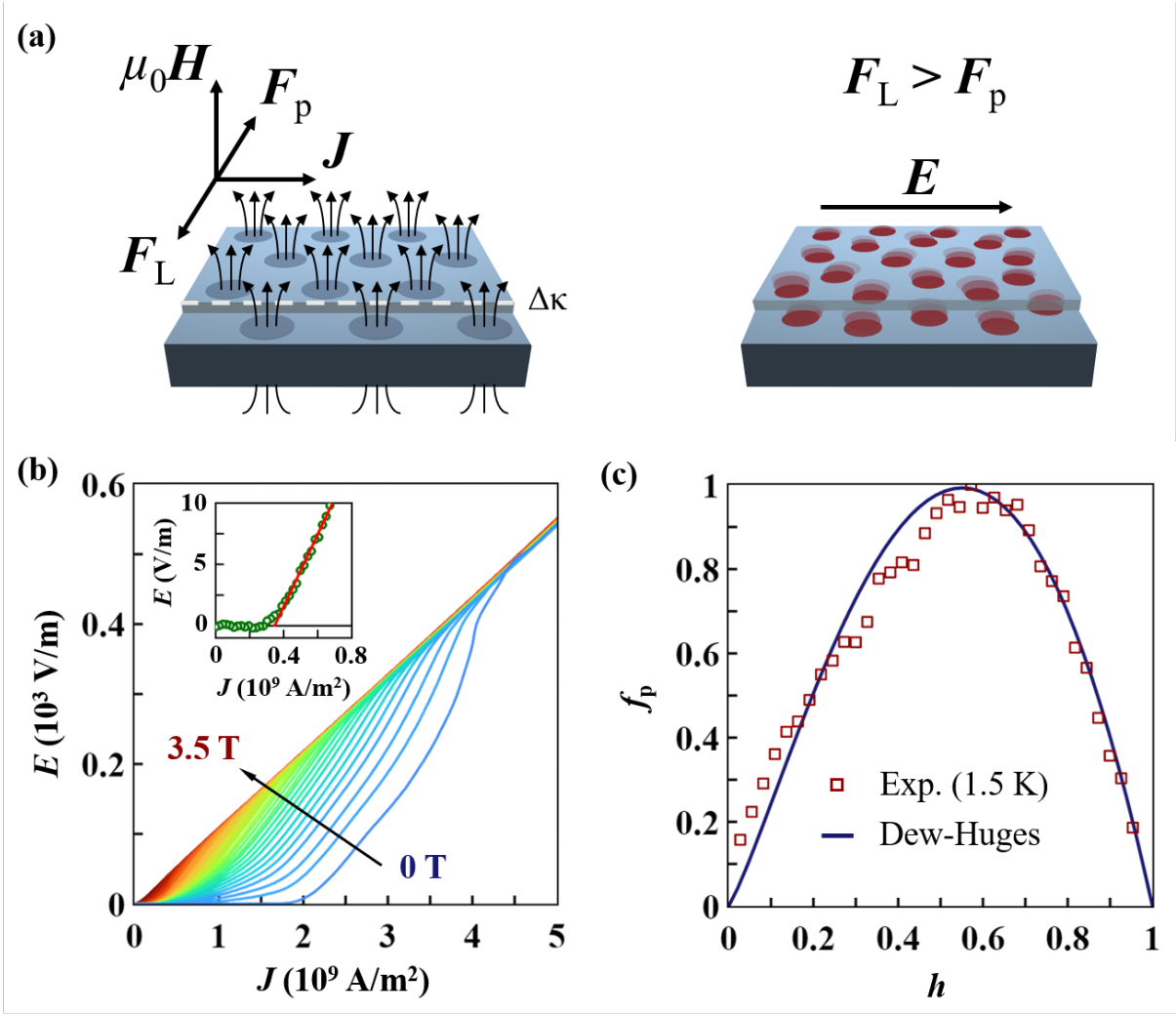

Figure 3. Characterization of flux pinning force in multilayer 2H-NbSe$_2$ step junction (a) Schematic image of vortices in a type-II superconductor step junction. Left: Forces acting on individual vortex. Right: Viscous flow of vortices. (b) Applied current density ($J$) dependence of Electric fields ($E$) under the out-of-plane magnetic fields ($B$) from 0.1 T to 3.5 T at 1.5 K. (Inset: $E$-$J$ characteristics at 1.5 K and B=0.6 T. The intersection of the $J$-axis with the linear fit (red line), gives the pinning current density) (c) The reduced flux pinning force ($f_p=F/F_p^{max}$) versus reduced out-of-plane magnetic fields ($h=B/B_p^{max}$) at 1.5 K. Red scatter is $f_p(h)$, and the blue line is the Dew-Hughes model fit.

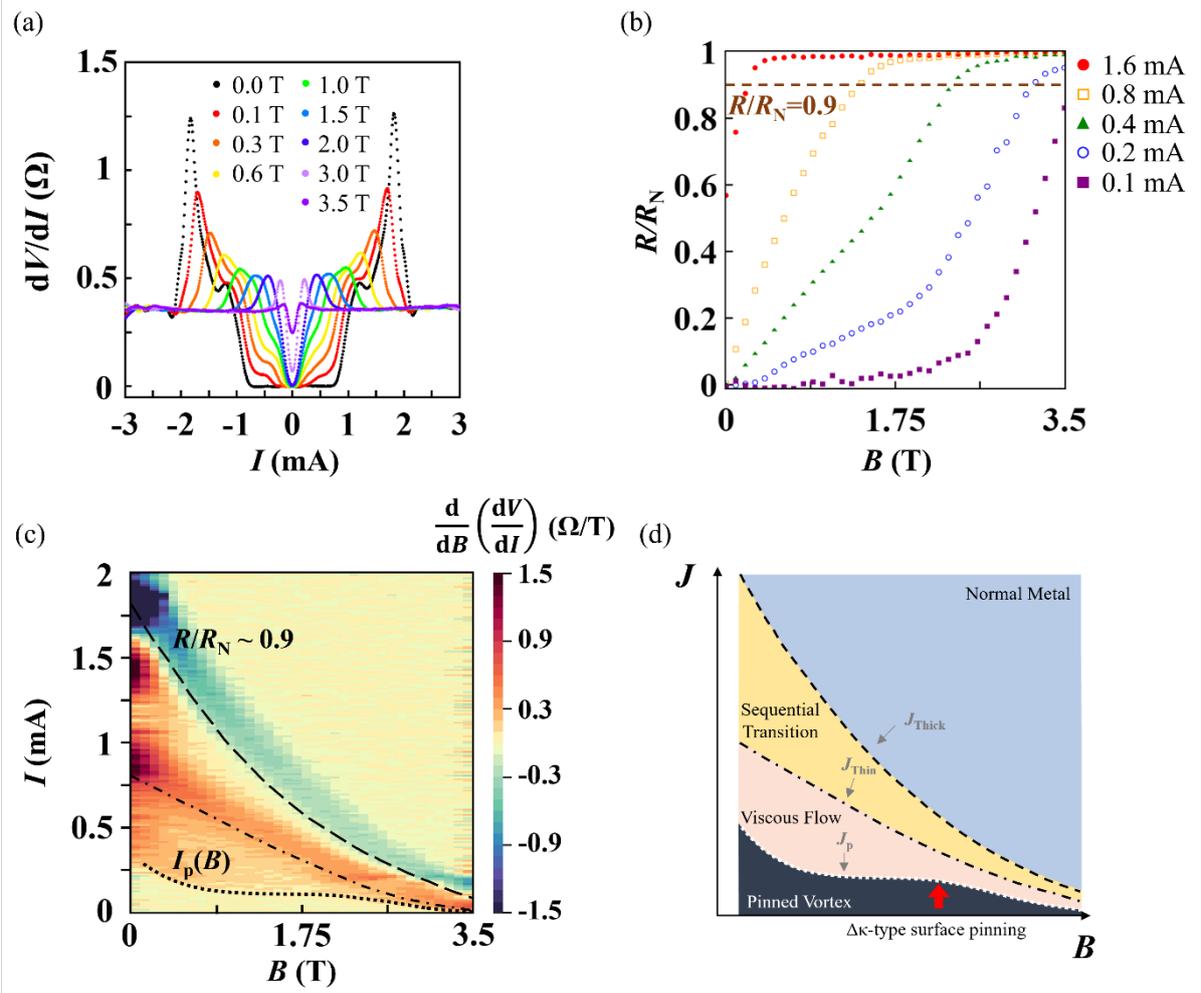

Figure 4. Superconducting phases in multilayer 2H-NbSe$_2$ step junction. (a) derivative of calculated differential resistance (d$V$/d$I$) with respect to $B$. (b) Out-of-plane magnetic fields ($B$) dependence of normalized resistance ($R$/$R_N$) at 1.5 K with various applied currents (c) Colourmap of d/d$B$(d$V$/d$I$) under $B$ from 0 T to 3.5 T at 1.5 K. Dashed line illustrates currents of $R$/$R_N \sim 0.9$ along $B$. The dotted line is the magnetic field-dependent pinning currents of $I_P$($B$). Dash-dot line is a guide to the eye. (d) Phase diagram of multilayer 2H-NbSe$_2$ step junction. The pinning current density ($J_p$) and the critical current densities ($J_{Thin}$ and $J_{Thick}$) separate the distinct superconducting phases. The red arrow indicates the point of maximum pinning force attributed to the Δκ-type surface pinning center.